\documentclass{sig-alternate-10pt}
\newfont{\mycrnotice}{ptmr8t at 7pt}
\newfont{\myconfname}{ptmri8t at 7pt}
\let\crnotice\mycrnotice%
\let\confname\myconfname%

\usepackage{ifpdf}
\usepackage{balance}
\usepackage{subfigure}
\usepackage{graphicx}
\usepackage{caption}
\usepackage[hyphens]{url}
\usepackage{url}
\usepackage[applemac]{inputenc}
\graphicspath{{artworks/}}

\usepackage[
  pass,
]{geometry}

\begin{document}

\title{ICONA: Inter Cluster ONOS Network Application}

\author{Matteo Gerola\textsuperscript{\dag}, 
Michele Santuari\textsuperscript{\dag}, 
Elio Salvadori\textsuperscript{\dag},
Stefano Salsano\textsuperscript{\S}, 
\and
Mauro Campanella\textsuperscript{\ddag},
Pier Luigi Ventre\textsuperscript{\ddag}, 
Ali Al-Shabibi\textsuperscript{$\ast$}, 
William Snow\textsuperscript{$\ast$}
\and
\affaddr{
\textsuperscript{\dag}CREATE-NET
\textsuperscript{\S}CNIT / Univ. of Rome Tor Vergata
\textsuperscript{\ddag}Consortium GARR
\textsuperscript{$\ast$}Open Networking Laboratory
}
}

\maketitle{}
\begin{abstract}
Several Network Operating Systems (NOS) have been proposed in the last few years for Software Defined Networks; however, a few of them are currently offering the resiliency, scalability and high availability required for production environments. Open Networking Operating System (ONOS) is an open source NOS, designed to be reliable and to scale up to thousands of managed devices. It supports multiple concurrent instances (a cluster of controllers) with distributed data stores. A tight requirement of ONOS is that all instances must be close enough to have negligible communication delays, which means they are typically installed within a single datacenter or a LAN network. However in certain wide area network scenarios, this constraint may limit the speed of responsiveness of the controller toward network events like failures or congested links, an important requirement from the point of view of a Service Provider. 
This paper presents ICONA, a tool developed on top of ONOS and designed in order to extend ONOS capability in network scenarios where there are stringent requirements in term of control plane responsiveness. In particular the paper describes the architecture behind ICONA and provides some initial evaluation obtained on a preliminary version of the tool.


\end{abstract}


\section{Introduction}
\label{section:introduction}

Since the beginning of the Software Defined Networking (SDN) revolution, the control plane reliability, scalability and availability were among the major concerns expressed by Service and Cloud Providers. Existing deployments show that standard IP/MPLS networks natively offer fast recovery in case of failures. Their main limitation lies in the complexity of the distributed control plane, implemented in the forwarding devices. IP/MPLS networks fall short when it comes to design and implementation of new services that require changes to the distributed control protocols and service logic. The SDN architecture, that splits data and control planes, simplifies the introduction of new services, moving the intelligence from the physical devices to a Network Operating System (NOS), also known as controller, that is in charge of all the forwarding decisions. The NOS is usually considered \textit{logically centralized} since it cannot introduce any single point of failure in production environments. Several \textit{distributed} NOS architectures have been proposed recently to guarantee the proper level of redundancy in the control plane: ONIX \cite{onix}, Kandoo \cite{kandoo}, HyperFlow \cite{hyperflow} to name a few.

One of the most promising solutions to properly deal with control plane robustness in SDN is ONOS (Open Networking Operating System) \cite{onos}. ONOS offers a stable implementation of a distributed NOS and it has been recently released under a liberal open-source license, supported by a growing community of vendors and operators. In ONOS architecture, a cluster of controllers shares a logically centralized network view: network resources are partitioned and controlled by different ONOS instances in the cluster and resilience to faults is guaranteed by design, with automatic  traffic rerouting in case of node or link failure. Despite the distributed architecture, ONOS is designed in order to control the data plane from a single location, even in case of large Wide Area Network scenarios.

However, as envisioned in the work of Heller et al. \cite{placement}, even though a single controller may suffice to guarantee round-trip latencies on the scale of typical mesh restoration target delays (200 msec), this may not be valid for all possible network topologies. Furthermore, ensuring an adequate level of fault tolerance can be guaranteed only if controllers are placed in different locations of the network.

To this purpose, we designed a geographically distributed multi-cluster solution called ICONA (Inter Cluster ONOS Network Application) which is working on top of ONOS and whose purpose is to  both increase the robustness to network faults by redounding ONOS clusters in several locations but also to decrease event-to-response delays in large scale networks.
In the rest of the paper we will consider a large scale Wide Area Network use-case under the same administrative domain (e.g. single service provider), managed by a single logical control plane. ICONA is a tool that can be deployed on top of ONOS; a first release is available under the Apache 2.0 open source license \cite {icona}.

The structure of the paper is as follows. Sect. \ref{section:onos} provides an overview of ONOS, while Sect. \ref{section:architecture} presents the ICONA architecture. Some preliminary results are then discussed in Sect. \ref{section:results}. Sect. \ref{section:sota} discusses the state-of-the-art in the field. Finally, Sect. \ref{section:conclusion} draws conclusions and indicates future works.

\section{An overview of {ONOS}}
\label{section:onos}

\begin{figure}
\centering
\includegraphics[width=.35\textwidth]{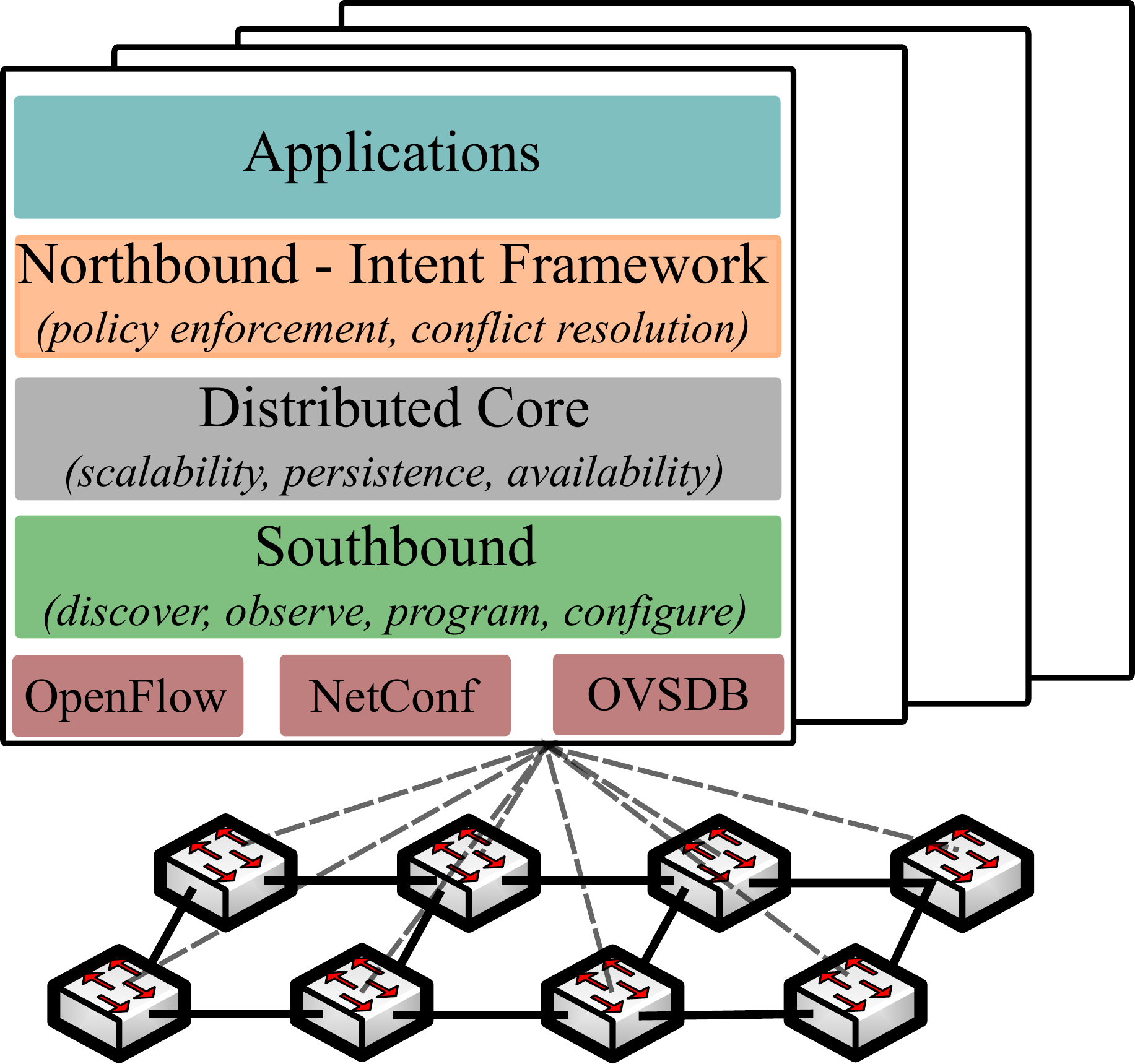}
\caption{ONOS distributed architecture}
\label{fig:onos}
\end{figure}

ONOS (Open Network Operating System) is a distributed SDN controller that has been recently released as open source by ON.Lab as part of a joint community effort with a number of partners including AT\&T, NEC and Ericsson among the others \cite{onos-web}. 

ONOS implements a distributed architecture in which multiple controller instances share multiple distributed data stores with eventual consistency. The entire data plane is managed simultaneously by the whole cluster. However, for each device a single controller acts as a master, while the others are ready to step in if a failure occurs. With these mechanisms in place, ONOS achieves scalability and resiliency. Figure \ref{fig:onos} shows the ONOS internal architecture within a cluster of four instances.
The Southbound modules manage the physical topology, react to network events and program/configure the devices leveraging on different protocols. The Distributed Core is responsible to maintain the distributed datastores, to elect the master controller for each network portion and to share information with the adjacent layers.
The NorthBound modules offer an abstraction of the network and the interface for application to interact and program the NOS. Finally, the Application layer offers a container in which third-party applications can be deployed.  

In case of a failure in the data plane (switch, link or port down), an ONOS instance detects the event (using the Southbound modules), computes alternative paths for all the traffic crossing the failed element (operation in charge of the Intent Framework), and publishes them on the Distributed Core: then, each master controller configures accordingly its network portion.

ONOS leverages on Apache Karaf \cite{karaf}, a java OSGi based runtime, which provides a container onto which various component can be deployed. Under Karaf, an application, such as ICONA, can be installed, upgraded, started and stopped at runtime, without interfering other components.

\section{ICONA architecture}
\label{section:architecture}

The basic idea behind ICONA is to partition the Service Provider's network into several geographical regions, each one managed by a different cluster of ONOS instances. While the management of the network is still under the same administrative domain, a peer-to-peer approach allows a more flexible design. The network architect can select the number of clusters and their geographical dimension depending on requirements (e.g. leveraging on some of the tools being suggested within the aforementioned work \cite{placement}), without loosing any feature offered by ONOS, neither worsening the system performances. In this scenario, each ONOS cluster provides both scalability and resiliency to a small geographical region, while several clusters use a publish-subscribe event manager to share topology information, monitor events and operator's requests.

\begin{figure}
\centering
\includegraphics[width=.45\textwidth]{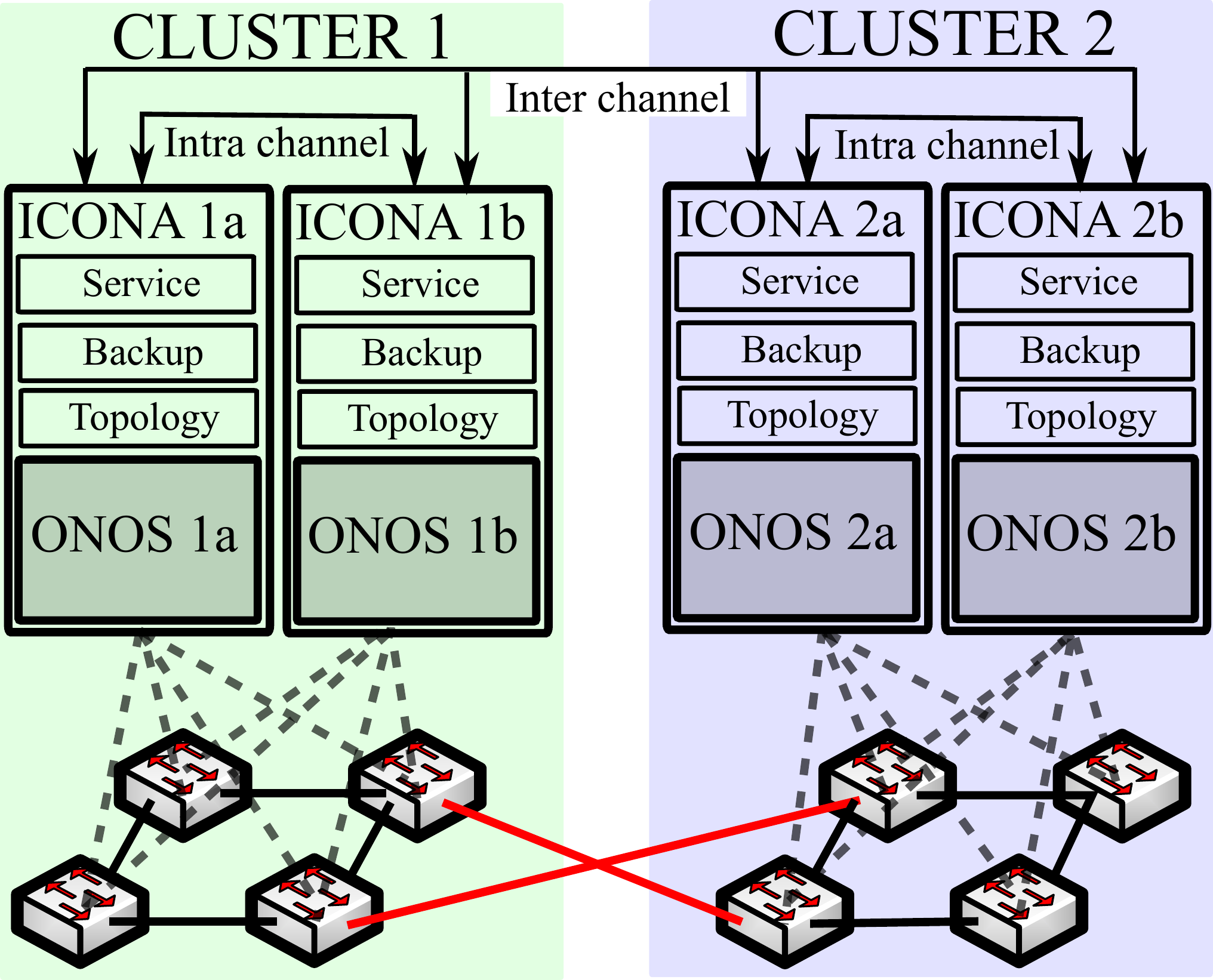}
\caption{ICONA architecture: the data plane is divided in portions interconnected through so-called inter-cluster links (in red). Each portion is managed by multiple co-located ONOS and ICONA instances, while different clusters are deployed in various data centers.}
\label{fig:arch}
\end{figure}

The ICONA control plane is geographically distributed among different data centers, each one controlling the adjacent portion of the physical network. Figure \ref{fig:arch} shows the aforementioned architecture where each cluster is located in a different data center (e.g. ONOS 1a and ONOS 1b are  co-located instances). To offer the same reliability already available in ONOS, the ICONA application runs on top of each instance. Referring on Figure \ref{fig:onos}, ICONA is deployed in the Application layer, and interacts with ONOS using the Northbound - Intent Framework.
 Inside each cluster, a master controller is elected, using a distributed registry. The ICONA master is in charge of sharing information and applying decisions, while all the backups are aware of the network status, and can become master in case of failure.

The communication between different ICONA instances, both intra and inter-cluster, is currently based on Hazelcast \cite{hazelcast}. Hazelcat offers a multicast event-based messaging system, inspired by Java Messaging System (JMS). Applications (e.g. controllers) can publish a message onto a topic, which will be distributed to all instances of the application that have subscribed to that topic. Hazelcast does not have a single point of failure that is not achieved easily by pure JMS solutions. In ICONA, one Hazelcast channel is devoted to the intra-cluster communication (e.g. local ONOS cluster) and another one for the inter-cluster messages.
ICONA runs as bundle in the ONOS Karaf container, taking advantages of all the  features mentioned in Sect. \ref{section:onos}. 

In devising a geographical architecture, covering thousands of square kilometers, a key element is the type and amount of information that the different segments of the control plane have to share.
Control traffic has to be minimized and the system has to optimize is performance in terms of:
\begin{itemize}
\item offering an up-to-date view of the network, including status of the nodes, 
\item configuring the network devices,
\item reacting to failures both in data and control plane without disrupting customer's traffic. 
\end{itemize}

Three internal modules implement these features in ICONA (see Figure \ref{fig:arch}): the Topology Manager, the Service Manager and the Backup Manager. In the following Sections an overview of each of these components is provided. 

\subsection{Topology Manager}

The Topology Manager (TM) is responsible to discover the data plane elements, reacts to network events, stores in a persistent local database the links and devices with the relevant metrics. To offer a partial view of its network portion, each TM shares with the other ICONA clusters the following information through the inter-cluster channel:
\begin{itemize}
\item Inter-cluster link (IL): an IL is the physical connection between two clusters. ICONA implements an enhanced version of the ONOS discovery mechanism, based on the Link Layer Discovery Protocol (LLDP). Each IL is shared with all the clusters tagged by some metrics, such as the link delay, available bandwidth and number of flows crossing the link.
\item End-point (EP) list: an EP defines the interconnection between the customer's gateway router and the ICONA network. Each cluster shares the list of its EPs and the metrics (bandwidth, delay) between these EPs and all the clusters ILs.
\end{itemize}

\subsection{Service Manager}

All the interaction between the network operators and the control plane are handled by the Service Manager (SM). This module offers a REST API used to poll ICONA for network events and alarms, and to manage (instantiate, modify and delete) the services. In our initial implementation, we have considered only two services: L2 pseudo-wire tunnels and MPLS VPN overlay networks, which are key services in a Service provider network. However, the application can be easily extended to provide other functionalities.

The implemented services require interconnecting two or multiple EPs that can be controlled by the same ICONA cluster or by different ones. The SM computes the overall metrics and chooses the best path between two EPs. There are two cases:

\begin{itemize}
\item If they belong to the same cluster, it directly asks ONOS to install an OpenFlow 1.3 \cite{openflow} MPLS-based flow path in the physical devices. 
\item If they belong to two different clusters, the SM follows this procedure:
\begin{itemize}
\item It contacts all the involved clusters (e.g. the ones that are crossed by this service request) asking to reserve the local portion of the path.
\item As soon as all the clusters have replied to the reservation event (if at least one does has a negative reply, the SM releases the path and computes an alternative ones), it requests to switch from the reservation status to the installation. Each ICONA cluster master then contacts the local ONOS asking to install the flow path in the physical devices.
\item If the second step is successful too, the SM ends the procedure, otherwise it asks all the clusters to release the resources, computes an alternative route and restarts the procedure.
\end{itemize}
\end{itemize}

Finally, the SM updates all the ICONA clusters with the new installed service and the relevant information.
While this path installation procedure is slower than the original implemented by ONOS (around three times), ICONA cannot assume a strong consistency between the remote databases. In the aforementioned ISP scenarios, the amount of time required for a service installation is not a relevant metric, but the procedure has to be robust enough to tolerate substantial delays.
  
\subsection{Backup Manager}

If a failure happens within the same cluster, ONOS takes care of rerouting all the paths involved in the failure. After receiving a \texttt{PORT\_STATUS} or \texttt{LINK\_DOWN} OpenFlow message, ONOS detects all the flows crossing the failed element and computes an alternative path to the destination. Finally, it installs the new flows in the network and removes the old ones.

If instead the failure happens between two different clusters (e.g. it involves an IL or one of the two nodes that sits at the edge of the IL) then the ICONA's module named Backup Manager (BM) will handle the failure. In particular:
\begin{itemize}
\item For each IL, a backup link (BIL) completely decoupled from the primary one, is computed and pre-installed in the data plane. The BIL is a virtual link selected among all the available paths between the source and destination ICONA clusters, taking into account composite metrics, such as the delay, available bandwidth and numbers of flows.
\item When the failure happens, all the traffic crossing the IL is rerouted to the BIL by each ICONA  edge cluster, without the need to wait for remote instances to share any information. 
\item When the rerouting is completed, the new path is notified to all the remote clusters, in order to share the same network status. 
\end{itemize}

The amount of time required for the rerouting is particularly relevant in an ISP network because it is most directly related to SLAs guaranteed by network operators.

\section{Evaluation}\label{section:results}

The purpose of the experimental tests described in this section is to compare ICONA with a standard ONOS setup, and evaluate the performances of the two solutions in an emulated environment.

The control plane is always composed of 8 virtual machines, each with four Intel Core i7-2600 CPUs @ 3.40GHz and 8GB of RAM. While for the ONOS tests all the instances belong to the same cluster, with ICONA we have created 2, 4 and 8 clusters, respectively with 4, 2 and 1 instances each. 
The data plane is emulated by Mininet \cite{mininet} and Netem \cite{netem}: the former creates and manages the OpenFlow 1.3 network, while the latter emulates the properties of wide area networks, such as variable delay, throughput and packet loss. Both solutions (ONOS and ICONA) have been tested with both a regular (grid) topology and with the topology of the G\'EANT \cite{geant} pan-European network. 
It is important to highlight that the current ONOS release (Avocet 1.0.1) is focusing on functions and correctness, while Blackbird, to be released in March, will dramatically improve the internal performance. For these reasons, the results presented in this section should not be considered as benchmark.


\subsection{Reaction to Network Events}
\subsubsection{Grid networks}

The first performance metric is the overall latency of the system for updating the network state in response to events; examples include rerouting traffic in response to link failure or moving traffic in response to congestion.
To evaluate how the system performs when the forwarding plane scales-out, we have selected some standard grid topologies, with a fixed link delay of 5ms (one-way) and then we have been comparing the latency needed to reroute a certain number of installed paths when an inter-cluster link fails for both ONOS and ICONA with various clustering settings.


 \begin{figure}
\centering
\includegraphics[width=.48\textwidth]{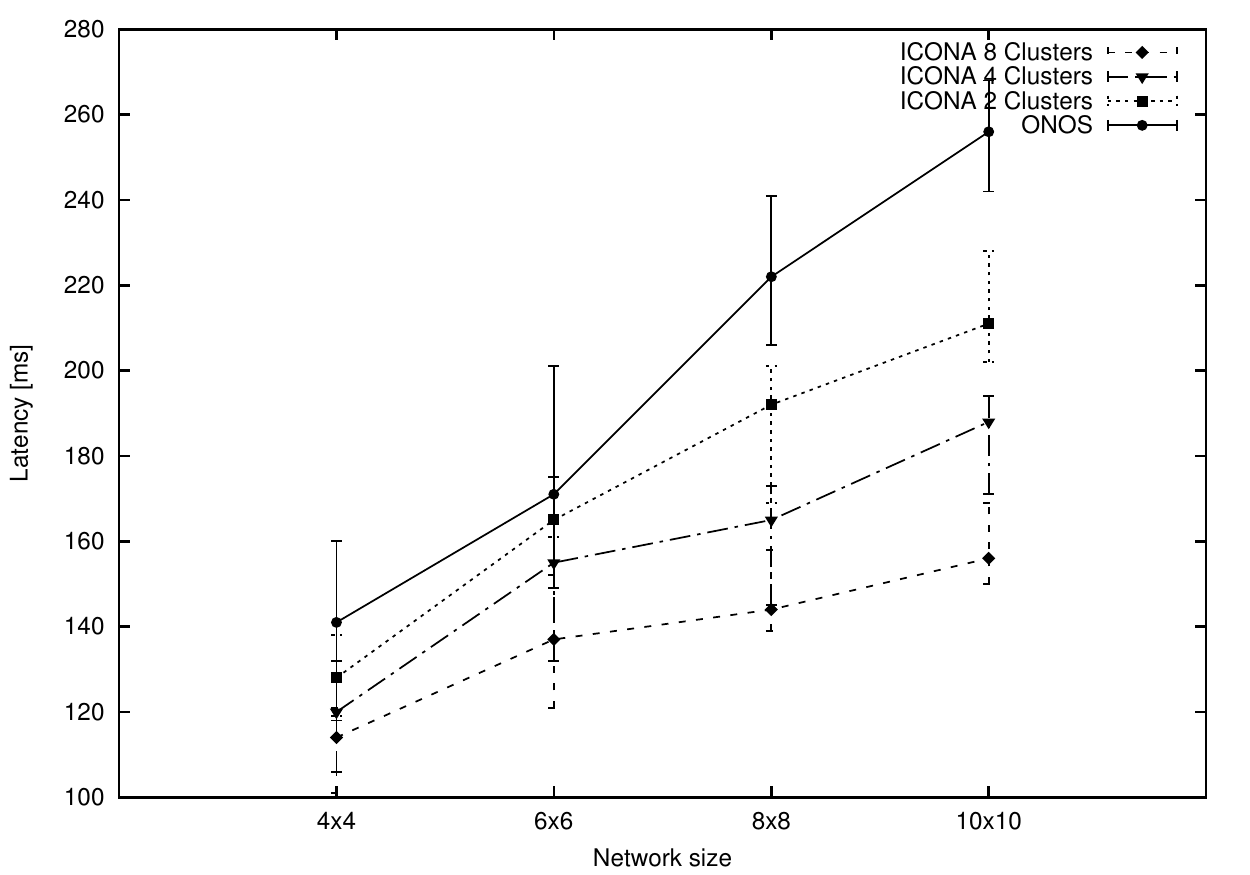}
\caption{Average, maximum and minimum latency to reroute 100 paths in case of link failure for ONOS and ICONA (2, 4 and 8 clusters)}
\label{fig:res_grid}
\end{figure}


We define as overall latency the amount of time that both ONOS and ICONA require to react to the failure as the sum of: i) the amount of time for the OpenFlow messages (\texttt{PORT\_STATUS} and \texttt{FLOW\_MOD}) to traverse the control network, ii) the alternative path computation, iii) the installation of new flows in the network devices and iv) the deletion of the pre-existing flows. 
In particular, we have been running several simulations by installing one thousand paths in the network and then by making fail an inter-cluster link with al least one hundred pseudo-wires running.

Figure \ref{fig:res_grid} shows the latency (avg, min, max) required for the different solutions to execute the four tasks previously mentioned. Each test has been repeated 20 times. Despite the same mechanism used by ICONA
to compute and install the new paths, the difference is mainly due to the following reasons: i) each ICONA cluster is closer to the devices, thus reducing the amount of time required for OpenFlow messages to cross the control channel and ii) the ICONA clusters are smaller, with fewer links and devices, thus decreasing the computation time and the overall numbers of flows to be installed and removed from the data plane.

\subsubsection{GEANT network}

\begin{figure}
\centering
\includegraphics[width=.40\textwidth]{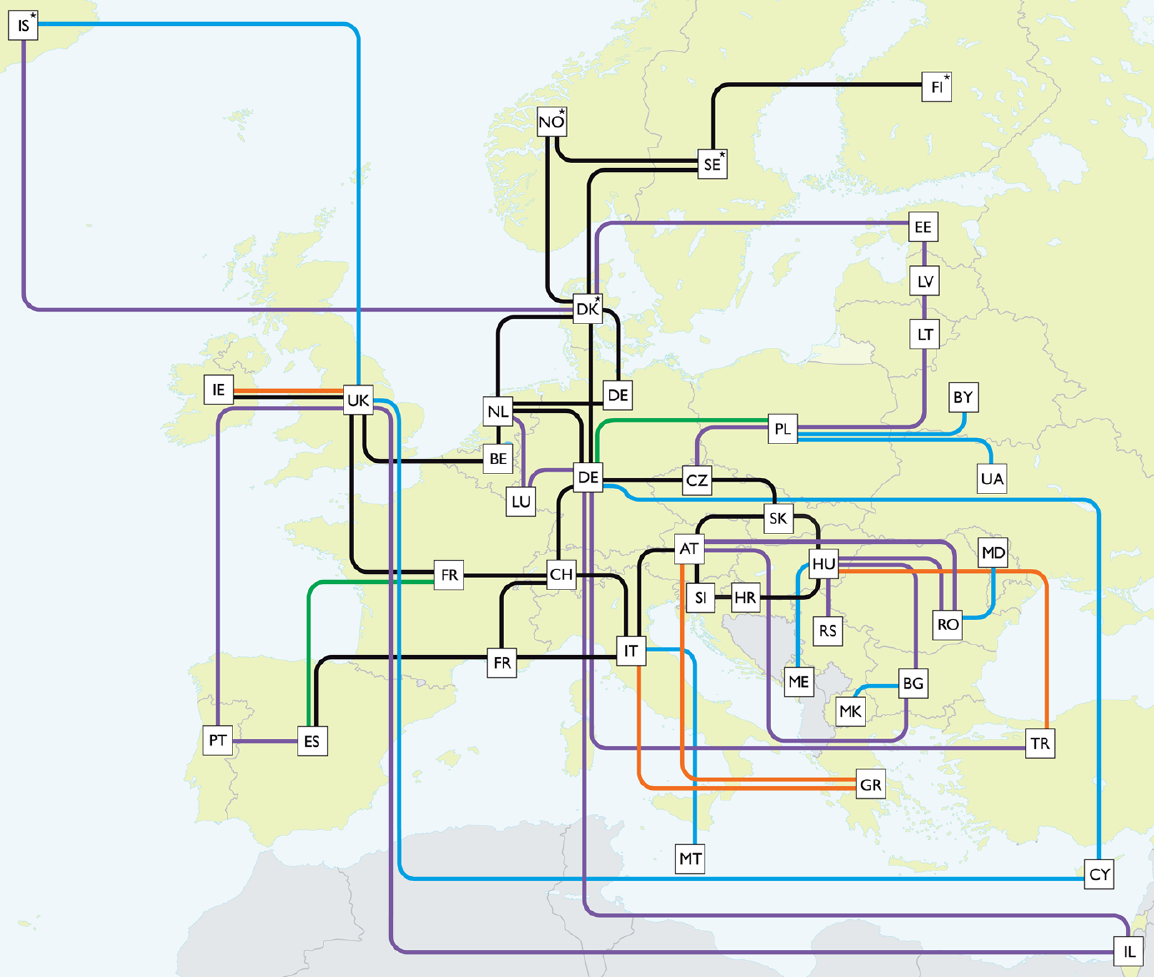}
\caption{G\'EANT pan-European network }
\label{fig:geant}
\end{figure}

The same metrics have been evaluated on the G\'EANT topology (see Figure \ref{fig:geant}). 
Circuits have various one-way delays (from 10 to 50ms) and throughputs (from 1 to 100Gbps). 
 
\begin{table}[h]
\begin{center}
\begin{tabular}{|p{1.5cm}|p{1.8cm}|p{1.8cm}|p{1.8cm}|}
\hline
\textbf{Control plane} & \textbf{Avg latency [ms]} & \textbf{Min latency [ms]} & \textbf{Max latency [ms]} \\ \hline
ONOS     &          297           & 284   & 308  \\ \hline
ICONA2         & 272          & 261 &   296  \\ \hline
ICONA4         & 246          & 232 &   257  \\ \hline
ICONA8           & 221          & 199 &   243  \\ \hline
\end{tabular}
\end{center}
\caption{G\'EANT network: average, maximum and minimum latency to reroute 100 paths in case of link failure for ONOS and ICONA (2, 4 and 8 clusters)}
\label{tab:geant_res}
\end{table}

Table \ref{tab:geant_res} depicts similar results as the previous test. 
ONOS has been compared with three different ICONA deployments, with 2, 4 and 8 clusters, respectively with 4, 2 and 1 instances each. While having the same number of VMs running the control plane software, their geographical distribution improves the overall performance of the system.
While the network is smaller than the 10*10 grid topology, with 41 switches and 58 bi-directional links, the higher delay in the data plane requires an additional amount of time to reconverge.

\subsection{Startup Convergence Interval}
This second experiment measures the overall amount of time required for both solutions to re-converge after a complete disconnection between the control and data planes. The tests have been performed over the G\'EANT topology, and replicated 20 times. Table \ref{tab:start_res} shows the average, maximum and minimum values in seconds. 

\begin{table}[h]
\begin{center}
\begin{tabular}{|p{1.5cm}|p{1.8cm}|p{1.8cm}|p{1.8cm}|}
\hline
\textbf{Control plane} & \textbf{Average Time [s]} & \textbf{Minimum Time [s]} & \textbf{Maximum Time [s]} \\ \hline
ONOS    &          6,98           & 6,95  & 7,06   \\ \hline
ICONA           &      6,96       &  6,88 & 7,02    \\ \hline
\end{tabular}
\end{center}
\caption{Amount of time required to obtain the network convergence after disconnection for ONOS and ICONA}
\label{tab:start_res}
\end{table}

The result shows that ICONA and ONOS require comparable time intervals to return to a stable state, in case of a complete shutdown or a failure of the control plane.

\section{Related Work}\label{section:sota}

The logical centralization of the control plane advocated by the SDN approach requires a specific design in terms of control network performance, scalability and fault tolerance. Most of the open source controllers currently available are focused on functionalities more than on scalability and fault tolerance. This section provides a review about distributed architectures for the SDN control plane, that address the scalability and fault tolerance issues.
ONIX \cite{onix} provides an environment on top of which a distributed NOS can be implemented with a logically centralized view. The distributed Network Information Base (NIB) stores the state of network in the form of a graph; the platform is responsible for managing the replication and distribution of the NIB, the applications have to detect and resolve conflicts of network state. Scalability is provided through network partitioning and aggregation. Regarding the fault tolerance, the platform provides the basic functions, while the control logic implemented on top of ONIX needs to handle the failures.
ONIX has been used in the B4 network, the private WAN \cite{jain} that inter-connects Google's data centers around the world. This high level design is similar in our ICONA solution, but ICONA is not tailored to a specific use case, providing a reusable framework on top of which it is possible to build specific applications.
The Kandoo \cite{kandoo} architecture addresses the scalability issue by creating an architecture with multiple controllers: the so-called root controller is logically centralized and maintains the global network state; the bottom layer is composed of local controllers in charge of managing a restricted number of switches. The Kandoo architecture does not focus on the distribution/replication of the root controller and on fault tolerance neither in the data plane nor in the control plane. HyperFlow \cite{hyperflow} focuses on both scalability and fault tolerance. Each HyperFlow instance manages a group of devices without losing the centralized network view. A control plane failure is managed by redirecting the switches to another HyperFlow instance. The applicability of such approach to WAN scenarios with large delays between the different HyperFlow instances is not considered.
DISCO \cite{disco} architecture considers a multi-domain environment. This approach is specifically designed to control a WAN environment, composed of different geographical controllers that exchange summary information about the local network topology and events. This solution overcomes the HyperFlow limitations, however it does not provide local redundancy: in the case of a controller failure, a remote instance takes control of the switches, increasing the latency between the devices and their primary controller.
ElastiCon \cite{elasticon} and Pratyaastha \cite{prat} aim to provide an elastic and efficient distributed SDN control plane to address the load imbalances due to static mapping between switches and controllers and spatial/temporal variations in the traffic patterns.
SMaRtLight \cite{smart} considers a distributed SDN controller aiming at a fault-tolerant control plane. It only focuses on control plane failures, assuming that data plane failures are dealt with by SDN applications on top of the control platform. 
The Beacon cluster \cite{beacon} project also targets the datacenter scenario, with a general framework focusing on load balancing (with addition and removal of controllers), dynamic switch-to-controller mapping and instance coordination.
In OpenDaylight \cite{odl}, an initial work on clustering has been provided in the last release of the project (Helium) using the Akka framework \cite{akka} and the RAFT consensus algorithm \cite{raft}.

\section{Conclusions}\label{section:conclusion}

In this paper we have presented ICONA (Inter Cluster ONOS Network Application), a tool built on top of ONOS distributed controller \cite{onos} whose aim is to enable clustering of ONOS instances across different locations on a Wide Area Network. In fact, while current ONOS release may scale and perform well on a variety of network topologies, it may not suffice for all, especially when there are latency bounds in restoration scenarios. ICONA is based on a pragmatic approach that inherits ONOS features while improving its responsiveness to network events like link/node failures or congested links that impose the rerouting of large set of traffic flows. 

A preliminary release of ICONA is available under permissive open source license and we are collaborating with ON.Lab to evaluate the opportunity to include some of the ideas developed therein as part of the official release of ONOS. Future improvements of ICONA will focus on solutions to improve the path installation procedure and on combining clustering design techniques with ICONA deployment in order to guarantee the best tradeoff between performances and replication of clusters in different network locations. 
\section{acknowledgements}
The authors thank the ONOS open source community for the continuous support and the software improvements.
This work was partly funded by the European Commission in the context of the DREAMER project, one of the beneciary projects of the G\'EANT Open Call research initiative.

\balance{}

\bibliographystyle{unsrt}
\bibliography{sosr2015}

\end{document}